\documentclass[aps,prd,twocolumn]{revtex4}
\usepackage{graphicx}

\newcommand{\bel}[1]{\begin{equation}\label{#1}}
\newcommand{\bal}[1]{\begin{eqnarray}\label{#1}}

\newcommand{\be}{\begin{equation}}
\newcommand{\ee}{\end{equation}}
\newcommand{\ba}{\begin{eqnarray}}
\newcommand{\ea}{\end{eqnarray}}

\newcommand{\bes}{\begin{equation*}}
\newcommand{\ees}{\end{equation*}}

\begin{document}

\title{Langevin dynamics of the pure $SU(2)$ deconfining transition}

\author{E. S. {\sc Fraga}$^1$\footnote{fraga@if.ufrj.br},
G. {\sc Krein}$^2$\footnote{gkrein@ift.unesp.br},
A. J. {\sc Mizher}$^1$\footnote{anajulia@if.ufrj.br} }

\affiliation{$^1$Instituto de F\'\i sica, Universidade Federal do Rio de Janeiro, \\
Caixa Postal 68528, 21941-972 Rio de Janeiro, RJ , Brazil \\
$^2$Instituto de F\'\i sica Te\'orica, Universidade Estadual Paulista, \\
Rua Pamplona 145, S\~ao Paulo, SP 01405-900, Brazil}

\begin{abstract}
We investigate the dissipative real-time evolution of the order parameter 
for the deconfining transition in the pure $SU(2)$ gauge theory. 
The approach to equilibrium after a quench to temperatures well above the 
critical one is described by a Langevin equation. To fix completely the markovian 
Langevin dynamics we choose the dissipation coefficient, that is a function of 
the temperature, guided by preliminary Monte Carlo 
simulations for various temperatures. Assuming a relationship 
between Monte Carlo time and real time, we estimate the delay in 
thermalization brought about by dissipation and noise.
\end{abstract}

\maketitle

\section{Introduction}

The study of the dynamics of phase conversion during the deconfinement 
transition for pure gauge theories might shed some light into the 
process of thermalization of the quark-gluon plasma in hot QCD in 
a controlled fashion. Indeed, for pure $SU(N)$, that can be seen as 
QCD in the limit of infinitely heavy quarks, lattice simulations are 
well developed, yielding a precise prediction for the deconfinement 
critical temperature and a good understanding of the corresponding 
thermodynamics \cite{Laermann:2003cv}. In this limit there is a global 
$Z(N)$ symmetry associated with the center of the gauge group, so that 
one can use the Polyakov loop to construct a well-defined exact order 
parameter \cite{Polyakov:1978vu,thooft}, and an effective 
Landau-Ginzburg field theory based on this quantity \cite{pisarski,ogilvie}. 

The effective potential for $T<<T_d$, where $T_d$ is the deconfinement 
critical temperature, has only one minimum, at the origin, 
where the whole system is localized. With the increase of the 
temperature, $N$ new minima appear. At the critical temperature, 
$T_d$, all the minima are degenerate, and above $T_d$ the new  
minima become the true vacuum states of the theory, so that the extremum 
at zero becomes unstable or metastable and the system 
starts to decay. In the case of $SU(3)$, whithin a range of temperatures 
close to $T_d$ there is a small barrier due to the weak first-order nature of 
the transition \cite{Bacilieri:1988yq}, and the process of phase 
conversion will thus be guided by bubble nucleation. For larger $T$, the 
barrier disappears and the system explodes in the process of spinodal 
decomposition \cite{reviews}. For $SU(2)$, the transition is 
second-order  \cite{Damgaard:1987wh}, and there is never a barrier to overcome. 

Real-time relaxation to equilibrium after a thermal quench followed by a 
phase transition, as considered above, can in general be described by standard 
reaction-diffusion equations \cite{reviews}. For a non-conserved order parameter, 
$\psi({\bf x},t)$, such as in the case of the deconfining transition in pure gauge theories, 
the evolution is given by the Langevin, or time-dependent Landau-Ginzburg, equation
\begin{equation}
\Gamma \, \frac{\partial\psi}{\partial t}=
-\frac{\delta F}{\delta\psi} + \xi =
-\sigma  \left[\frac{\partial^2\psi}{\partial t^2} - \nabla^2 \psi 
\right] - \frac{\partial U}{\partial\psi}  +  \xi \; ,
\label{TDLG}
\end{equation}
where $F=F(\psi,T)$ is the coarse-grained free energy of the system, 
$\sigma$ is the surface tension and $U=U(\psi,T)$ is the effective potential. 
The quantity $\Gamma$ is known as the dissipation coefficient and will play 
an important role in our discussion. Its inverse defines a time scale for 
the system, and is usually taken to be either constant 
or as a function of temperature only, $\Gamma=\Gamma (T)$. The function $\xi$ 
is a stochastic noise assumed to be gaussian and white, so that 
\begin{eqnarray}
\langle\xi ({\bf x}, t)\rangle&=&0 \, , \nonumber \\
\langle \xi ({\bf x}, t)\xi({\bf x}' ,t')\rangle&=&
2\Gamma \delta ({\bf x}- {\bf x}' )\delta (t - t') \, , 
\end{eqnarray}
according to the fluctuation-dissipation theorem. 
{}From a microscopic point of view, the noise and dissipation terms 
are originated from thermal and quantum fluctuations resulting 
either from self-interactions of the field representing the order parameter or 
from the coupling of the order parameter to different fields in the 
system. In general, though, Langevin 
equations derived from a microscopic field theory \cite{micro} contain also the 
influence of multiplicative noise and memory 
kernels \cite{FKR,Koide:2006vf,FKKMP}. 

In this paper, we consider the pure $SU(2)$ gauge theory, 
without dynamical quarks, that is rapidly driven to very high 
temperatures, well above $T_d$, and decays to the deconfined phase via 
spinodal decomposition. We are particularly interested in the effect of 
noise and dissipation on the time scales involved in this ``decay 
process'', since this might provide some insight into the problem of 
thermalization of the quark-gluon plasma presumably formed in high-energy 
heavy ion collisions \cite{bnl}. For the order parameter and effective potential we 
adopt the effective model proposed in Ref. \cite{ogilvie}, and the choice of 
the dissipation coefficient, that is a function of the temperature, is guided by 
preliminary Monte Carlo simulations for various temperatures, 
comparing the short-time exponential growth of the two-point Polyakov 
loop correlation function predicted by the simulations \cite{Krein:2005wh} to the 
Langevin description assuming, of course, that both dynamics are the same 
(see, also, the extensive studies of Glauber evolution by Berg {\it et al.} \cite{berg}). 
This procedure fixes completely the Markovian Langevin dynamics, as will be 
described below, if one assumes a relationship between Monte Carlo time and real 
time. Once the setup is defined for the real-time evolution, we can estimate the delay 
in thermalization brought about by dissipation and noise by performing numerical 
calculations for the dynamics of the order parameter on a cubic lattice. As will 
be shown in the following, the effects of dissipation and noise significantly delay 
the thermalization process for any physical choice of the parameters, a result 
that is in line with but is even more remarkable than the one found for the chiral 
transition \cite{Fraga:2004hp}.

The paper is organized as follows. In Section II, we describe the effective model 
adopted for the Langevin evolution implementation, as well as the analytic behavior 
for early times. In Section III, we discuss 
the necessity of performing a lattice renormalization to have results that thermalize 
to values that are independent of the lattice spacing and free from ultraviolet 
divergences, and present the necessary counterterms. 
In Section IV we briefly describe the Glauber dynamics of pure lattice gauge theory 
that can be used to extract the dissipation coefficient for different values of the 
temperature. Details and quantitative results from lattice simulations will be 
presented in a future publication \cite{next}. In Section V we present and 
analyze our numerical results for the time evolution of the order parameter for 
deconfinement after a quench to temperatures above $T_d$. Finally, Section 
VI contains our conclusions and outlook.

\section{Effective model and Langevin dynamics}

Since we focus our investigation on pure gauge $SU(N)$ theories, we can adopt 
effective models built by using functions of the Polyakov loop as the order parameter 
for the deconfining phase transition. If quarks were included in the theory, the $Z(N)$ 
symmetry present in pure glue systems would be explicitly broken, and the Polyakov 
loop would provide only an approximate order parameter. For euclidean gauge theories 
at finite temperature, one defines the Polyakov loop as:
\begin{equation}
P(\vec x) = \mathcal{T} \exp \left[ig\int^{1/T}_0 d\tau A_0(\vec x, \tau)\right]  \, ,
\end{equation}
where $\mathcal{T}$ stands for euclidean time ordering, $g$ is the gauge coupling constant 
and $A_0$ is the time component of the vector potential. 

The effective theory we adopt \cite{ogilvie} is based on a mean-field treatment in which 
the Polyakov loops are constant throughout the space. The degrees of freedom that will 
be used to construct the free energy are the eigenvalues of the Polyakov loop, rather 
than $\langle Tr_F P(\vec x)\rangle$. Working in $SU(N)$ gauge theories the Polyakov 
loop is unitary, so that it can be diagonalized by a unitary transformation, assuming the 
form
\begin{equation}
P_{jk} = \exp(i\theta_j) \; \delta_{jk}  \, .
\label{parametrized_loop}
\end{equation}

At one loop, the free energy for gluons in a constant $A_0$ background is given by
\begin{equation}
f_{pert}(\theta) = ln [det(-D^2_{adj})]  \, ,
\end{equation}
where $D_{adj}$ is the covariant derivative acting on fields in the adjoint representation. 
This expression can be written in a more explicit form:
\begin{eqnarray}
f &=& -\frac{1}{\beta}\sum^{N}_{j,k=1}2\left(1-\frac{1}{N}\delta_{jk}\right)
\times \nonumber \\
&&\int\frac{d^3 k}{(2\pi)^3}\sum_{n=1}^{\infty}\frac{1}{n} e^{- n\beta\omega_k +
in \Delta\theta_{jk}}  \, ,
\label{f_perturbative}
\end{eqnarray}
where $\theta$ is defined in Eq.(\ref{parametrized_loop}), and 
$\Delta\theta_{jk}\equiv \theta_j-\theta_k$. Here we have the ``bare'' dispersion relation 
$\omega_{\bf k}\equiv|\bf k|$. In order to include confinement in this effective model 
description, one can introduce an {\it ad hoc} ``thermal mass'' for the gluons, so 
that the dispersion relation becomes $\omega_{\bf k}=\sqrt{{\bf k}^2 + M^2}$. The 
value of $M$ can be related to the critical temperature $T_d$ extracted from 
lattice simulations.

\begin{figure}[htb]
\includegraphics[width=8cm]{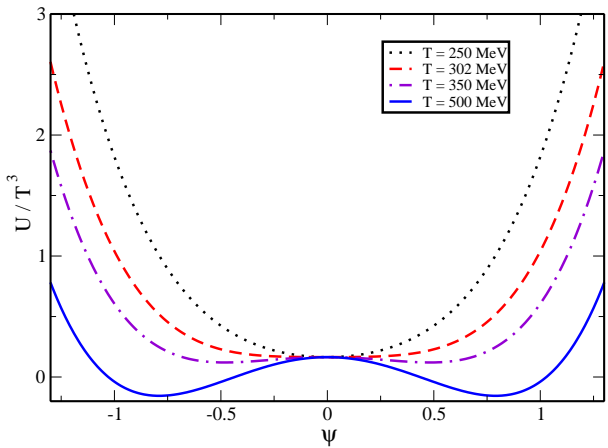}
\caption{Effective potential for the case of $SU(2)$ for different values of the temperature.} 
\label{pot_su2}
\end{figure}

Parametrizing the diagonalized Polyakov loop as 
$diag[\exp(i\phi_{N/2},...,i\phi_1,-i\phi_1,...,-i\phi_{N/2}]$, we can construct the 
effective potential from the free energy above. For $SU(2)$, it can be written in 
the following convenient form:
\begin{eqnarray}
\frac{U}{T^3} &=& \frac{\pi^2}{60} - \frac{\pi^{2}}{6}\left( 1-\frac{T_d^2}{T^2}  \right)\psi^2 
+ \frac{\pi^{2}}{12}\psi^4   \; ,
\end{eqnarray}
where we have defined  $\psi\equiv 1-2\phi/\pi$, and used the relation between the 
mass $M$ and the critical temperature $T_d=\sqrt{3M^2/2\pi^2}$. In Fig. \ref{pot_su2} 
we display $U$ as a function of $\psi$ for different values of the temperature. 
One can see from this plot that for $T\ll T_d$ the minimum is at $\psi({\bf x}, t)=0$. 
As the temperature increases new minima appear;  above the critical temperature 
they become the true vacuum states of the system. Now, if at $t=0$ the temperature is 
rapidly increased to $T \gg T_d$, the system is brought to an unstable state and 
therefore will start ``rolling down" to the new minima of the effective potential. 
%
%

To study the time evolution, we consider a system characterized by a coarse-grained 
free energy of the form
\begin{equation}
F(\psi, T) = \int d^3x\left[\frac{\sigma}{2}(\nabla\psi)^2 + U(\psi, T)\right]  \, ,
\end{equation}
where $U$ is the effective potential obtained above, and $\sigma$ plays the role of a 
surface tension \cite{Bhattacharya:1990hk}, assuming the following value for $SU(2)$: 
$\sigma=\pi^2T/g^2$, $g$ being the gauge coupling. 
The approach to equilibrium of the order parameter $\psi({\bf x},t)$ will then be dictated 
by the Langevin equation (\ref{TDLG}) that, for arbitrary times, has to be solved numerically 
on a lattice. 

At very short times, however, when $\psi(\mathbf{x},t) \approx 0$, non-linear terms in the 
evolution equation can be neglected, so that Eq. (\ref{TDLG}) reduces to
\begin{equation}
\sigma \ddot\psi_{\bf k} + \Gamma\dot\psi_{\bf k} + \sigma{\bf k}^2\psi_{\bf k} - 
T^3 m_T^2 \psi_{\bf k} \approx 0
\label{linear-TDLG}
\end{equation}
in the Fourier space, where $m_T$ is a dimensionless thermal mass that can be written as 
\begin{equation}
m_T^2=\frac{\pi^{2}}{3} \epsilon(T) = 
\frac{\pi^{2}}{3} \left( 1 - b ~\frac{T_d^2}{T^2}  \right) \, .
\end{equation}
$b$ is a number that depends on the details of the quadratic term of the particular 
effective potential adopted, so that it will be different, for instance, if we consider 
$SU(2)$ ($b=1$) or $SU(3)$ ($b=10/9$). One can, then, approximate the (noiseless) 
solution in Fourier space by $\psi({\bf k}, t \approx 0) \sim e^{\lambda_{\bf k} t}$, 
where $\lambda_{\bf k}$ 
are the roots of the quadratic equation
\begin{equation}
\lambda_{\bf k}^2 + \left( \frac{\Gamma}{\sigma} \right)\lambda_{\bf k} 
+ \left( {\bf k}^2 - \frac{T^3 m_T^2}{\sigma} \right) = 0 \, .
\end{equation}
For wavelength modes such that 
\begin{equation}
|{\bf k}| <  \left[ \left(\frac{\Gamma}{2\sigma}\right)^2 + 
\left( \frac{T^3 m_T^2}{\sigma} \right) \right]^{1/2} \, ,
\end{equation}
one has already oscillations, but those are still damped by a factor 
$\exp(-\Gamma t/ 2\sigma)$. It is only for longer wavelength modes, i.e. 
\begin{equation}
|{\bf k}| < k_c = \left(\frac{T^3 m_T^2}{\sigma}\right)^{1/2} \, , 
\end{equation}
that there will be an explosive exponential growth corresponding to the 
regime of spinodal decomposition. 

As time increases, the order parameter increases and non-linear contributions 
take over. To study the complete evolution of the phase conversion process, we 
have to solve $(\ref{TDLG})$ numerically on a lattice. In the next section we discuss 
the need for lattice renormalization to avoid spurious ultraviolet divergences in 
the dynamics.

\section{Lattice renormalization}

In performing lattice simulations of the Langevin evolution, one should 
be careful in preserving the lattice-size independence of the results, especially 
when one is concerned about the behavior of the system in the continuum limit. 
In fact, in the presence of thermal noise, short and long wavelength modes are 
mixed during the dynamics, yielding an unphysical lattice size sensitivity. 
The issue of obtaining robust results, as well as the correct ultraviolet 
behavior, in performing Langevin dynamics was discussed by several 
authors \cite{Borrill:1996uq,Bettencourt:1999kk,Gagne:1999nh,
Bettencourt:2000bp,krishna}. The problem, which is not {\it a priori} evident 
in the Langevin formulation, is related to the well-known Rayleigh-Jeans 
ultraviolet catastrophe in classical field theory. The dynamics dictated by 
Eq. (\ref{TDLG}) is classical, and is ill-defined for very large momenta. 

Equilibrium solutions of the Langevin equation that are 
insensitive to lattice spacing can be obtained, in practice, by adding 
finite-temperature counterterms to the original effective potential, 
which guarantees the correct short-wavelength behavior of the discrete theory. 
Furthermore, it assures that the system will evolve to the correct quantum state 
at large times. For a more detailed analysis of lattice renormalization 
in the Langevin evolution, including the case of multiplicative noise, 
see Ref. \cite{noise-broken}.

Since the classical scalar theory in three spatial dimensions is 
super-renormalizable, only two Feynman diagrams are divergent, the tadpole 
and sunset. The singular part of these graphs can be isolated using lattice 
regularization, and then subtracted from the effective potential in the Langevin 
equation. For a scalar field theory, explicit expressions for the counterterms 
were obtained by {\it Farakos et al.}  \cite{Farakos:1994kx} within the framework 
of dimensional reduction in a different context.

Following Ref. \cite{Farakos:1994kx}, we write the bare potential 
in a three-dimensional field theory in the following form
\begin{equation}
{\cal V}(\phi) = - \frac{1}{2} m^2 \phi^2 + \frac{1}{4} \lambda_3 \phi^4 \, ,
\label{pot-3d}
\end{equation}
where $m$ is the bare mass of the field $\phi$ and the subindex in $\lambda_3$ 
stresses the fact that this is the coupling of a three-dimensional theory. 
In Ref. \cite{Farakos:1994kx}, this dimensionally-reduced theory was obtained 
from a four-dimensional theory with a dimensionless coupling $\lambda$, 
assuming a regime of very high temperature. At leading order, one has 
$\lambda_3 = \lambda \, T$.  The mass counterterm, which is defined such that
\begin{equation}
- \frac{1}{2} \, m^2 \, \phi^2  \rightarrow  -  
\frac{1}{2} \, \left( m^2 + \delta m^2 \right)  \, \phi^2  \equiv
- \frac{1}{2} \, m^2_R  \, \phi^2
\end{equation}
is given by
\begin{equation}
\delta m^2 = 3 \,  \lambda_ 3 \; \frac{0.252731}{a} - 6 \, \lambda^2_3 \;
\frac{1}{16\pi^2} \left[\ln\left(\frac{6}{\mu a}\right) + 0.09 \right]   \, ,
\end{equation}
where $a$ is the lattice spacing and $\mu$ is the renormalization scale. The first 
term comes from the tadpode diagram and the second one from the sunset. 
Finite constants are obtained imposing that, after renormalization, the sunset 
diagram yields the same value for three renormalization schemes: lattice, 
momentum subtraction and $\overline{MS}$ \cite{Farakos:1994kx}. 
Notice that in order to obtain lattice-independent results physical
quantities become $\mu$-dependent~\cite{Gagne:1999nh}. 
However, since the contribution from the $\mu$-dependent term is logarithmic, 
variations around a given choice for this scale affect the final results by a 
numerically negligible factor, as we verified in our simulations, so that 
this dependence is very mild.

Since the field $\psi$ in the effective model we consider here is dimensionsless, 
it is convenient to define the dimensionful field $\varphi = \sigma^{1/2} \psi$ 
in order to relate results from Ref. \cite{Farakos:1994kx} to our case more 
directly.

Now we can write our Langevin equation, Eq.~(\ref{TDLG}), in terms  of the field 
$\varphi$. For $SU(2)$, we have
\begin{equation}
\left( \frac{\partial^2 \varphi}{\partial t^2}  - \nabla^2 \varphi
\right) +  \frac{\Gamma}{\sigma} \frac{\partial \varphi}{\partial t} - m^2_L \, \varphi
+ \lambda_L \varphi^3   = \frac{\xi}{\sigma^{1/2}}  \, 
\label{TDLG-varphi}
\end{equation}
where 
\begin{eqnarray}
m^2_L &=& \frac{T^{3} m_T^2}{\sigma}= \frac{\epsilon(T)}{3} \, g^2 \, T^2  \, , \\
\lambda_L &=&   \frac{\pi^{2} T^3}{3 \sigma^2}=\frac{1}{3\pi^2} \, g^4 \, T  \, .
\end{eqnarray}
The subindex $L$ in these quantities is a reminder that they refer to the Langevin
equation. It is clear that Eq.~(\ref{TDLG-varphi}) corresponds to an effective action 
${\cal S}(\varphi)$ given by
\begin{equation}
{\cal S}(\varphi) =  \frac{1}{2} (\nabla\varphi)^2 - \frac{1}{2} m^2_L \,  \varphi^2
+ \frac{1}{4} \lambda_L \,  \varphi^4 \, .
\end{equation}

Once we have identified the mass term and the coupling constant, we can renormalize
the Langevin equation, which becomes
\begin{eqnarray}
\left( \frac{\partial^2 \varphi}{\partial t^2}  - \nabla^2 \varphi
\right) +  \frac{\Gamma}{\sigma} \frac{\partial \varphi}{\partial t} 
&=& (m^2_L + \delta m^2_L )  \, \varphi  \nonumber \\
&-& \lambda_L \varphi^3   + \frac{\xi}{\sigma^{1/2}}  \, ,
\end{eqnarray}
where
\begin{equation}
\delta m^2_L =  3 \,  \lambda_ L \; \frac{0.252731}{a} - 6 \, \lambda^2_L \;
\frac{1}{16\pi^2} \left[\ln\left(\frac{6}{\mu a}\right) + 0.09 \right]  \, ..
\end{equation}
Notice that we have used the same symbol $\varphi$ to denote both the renormalized 
and non-renormalized fields, since the theory is super-renormalizable and only mass 
counterterms are needed. In terms of the original $\psi$, our renormalized Langevin 
equation is finally given by
\begin{eqnarray}
\frac{\pi^2 T}{g^2} \left( \frac{\partial^2 \psi}{\partial t^2}  - \nabla^2 \psi
\right)  +  \Gamma \frac{\partial \psi}{\partial t} &=& \left[ \frac{\pi^2 T^3}{3} \epsilon(T)
+ \delta {\cal  M}^2_\psi \right] \psi   \nonumber \\
&-&  \frac{\pi^2 T^3}{3} \psi^3   + \zeta  \, ,
\label{GLL-renorm}
\end{eqnarray}
where
\begin{equation}
\delta {\cal M}^2_\psi = \frac{\pi^2 T}{g^2} \; \delta m^2_L  \, .
\end{equation}
One can factor out the appropriate powers of $T$ in this expression 
to make explicit the mass dimensions:
\begin{eqnarray}
\delta {\cal M}^2_\psi &=& 3 \, T^2  \, \frac{0.252731}{a} \left( \frac{g^2}{3} \right)
\nonumber \\
&-& 6 \, T^3 \,  \frac{1}{16\pi^2} \left[\ln\left(\frac{6}{\mu a}\right) + 0.09 \right]
\left( \frac{g^2}{3} \right)^2  \left( \frac{g^2}{\pi} \right) . \nonumber \\
\end{eqnarray}

Notice that for sufficiently high temperatures the 
symmetry of the potential is restored. 

\section{Dissipation coefficient from Monte Carlo evolution}

In lattice simulations for pure $SU(N)$ gauge theories, one 
can implement the Glauber dynamics by starting from thermalized gauge field 
configurations at a temperature $T < T_d$ and then changing the temperature of 
the entire lattice that is quenched to $T >T_d$ \cite{berg,next}. 
The gauge fields are then updated using the heat-bath
algorithm of Ref.~\cite{updating} without over-relaxation. A ``time'' unit 
in this evolution is defined as one update of the entire lattice by
visiting systematically one site at a time. 

The structure function, defined as
\begin{equation}
S(k,\tau) = \langle \widetilde L_F(k,\tau)  \widetilde L_F(-k,\tau) \rangle \, ,
\end{equation}
where $\widetilde L_F(k,\tau)$ is the Fourier transform of
$L_F(x,\tau)$, the Polyakov loop in the fundamental representation, 
can be used to obtain the values of the dissipation coefficient, $\Gamma$, 
for different values of the final temperature, $T$, as follows. At early times, 
immediately after the quench, $\psi \simeq 0$ and one can neglect the terms
proportional to $\psi^3$ and $\psi^4$ in the effective potential to first approximation. 
It is not difficult to show that at early times, when $\psi$ is small, the
structure function can be written as
\begin{equation}
S(k,\tau)= S(k,0) \, \exp\left[ 2\omega(k) \, \tau\right] \, ,
\end{equation}
where
\begin{equation}
\omega(k) = \frac{\pi^2 T}{g^2 \Gamma}\; \left(k^2_c - k^2\right) \, .
\end{equation}
In obtaining this expression we have neglected the second-order 
time derivative in Eq. (\ref{linear-TDLG}), which should be a good 
approximation for a rough estimate of $\Gamma$. 
For the effective potential adopted here, $k^2_c$ is given by
\begin{equation}
 k^2_c =\frac{g^2}{3}\left(T^2 - \frac{9 M^2}{4\pi^2}\right) .
\end{equation}
One sees that for momenta smaller than the critical momentum
$k_c$, one has the familiar exponential growth, signaling
spinodal decomposition. Plotting $ \ln S(k,\tau)/\tau$ for
different values of $k$ allows one to extract $2\omega(k)$ and, in
particular, the value of $k^2_c$. Once one has extracted these
values, $\Gamma$ can be obtained from the following relation:
\begin{equation}
\Gamma^{-1} =  \omega(k) \, \frac{g^2}{\pi^2 T(k^2 - k^2_c)} \, .
\end{equation}

Now, in Monte Carlo simulations one does not have a time variable
in physical units and so, by plotting $\ln S$ from the lattice, one
obtains values of $2\omega(k)$ that do not include the (unknown) scale
connecting real time $\tau$ and Monte Carlo time. Nevertheless, if one
assumes that the relation between the Langevin time variable
$\tau$ and the Monte Carlo time is linear, one can parametrize
this relation in terms of the lattice spacing $a$ as $\tau = a\lambda_{MC}$, 
where $\lambda_{MC}$ is a dimensionless parameter
that gives this relation in units of the lattice spacing. 
An estimate for the relationship between Monte Carlo time and real time 
is given in Ref.~\cite{Tomboulis:2005es}, where the authors evaluate 
the number of sweeps necessary for the system to freeze-out. In 
this reference, the authors implement lattice Monte Carlo simulations 
of the change of the Polyakov loop under lattice expansion and 
temperature falloff. The freeze-out number of sweeps was
defined as being the number of sweeps necessary for the Polyakov
loop to reach zero. This number was found to be of the order of
$5000$ for the range of temperatures we are considering here. Using
the phenomenological value of $9$~fm/c~\cite{Kolb:2003gq} as the
freeze-out time, one can then obtain $\lambda_{MC}$.

Preliminary simulations clearly show that $\Gamma^{-1}$ decreases as 
the final temperature increases \cite{next}. Guided by these 
results, we choose in the case of $SU(2)$ 
$\Gamma (T)=10^3~$fm$^{-2}$ for our Langevin simulations, which we 
describe in the next section.

\section{Numerical results for deconfinement and discussion}

We solve Eq. (\ref{TDLG}) numericaly for $SU(2)$ in a cubic spacelike 
lattice with $64^3$ sites under periodic boundary conditions, using the semi-implicit 
finite-difference method for time discretization and finite-difference Fast Fourier Transform 
for spatial discretization and evolution \cite{copetti}. To compute the expectation value 
of the order parameter $\psi$, we average over thousands of realizations with different 
initial conditions around $\psi\approx 0$ and different initial configurations for the noise. 
At each time step we compute
\begin{equation}
\langle \psi\rangle = \frac{1}{N^3}\sum_{ijk}\psi_{ijk} (t) \, ,
\end{equation}
where the indices $i,j,k$ indicate the position of the site on the lattice. 

\begin{figure}[htb]
\includegraphics[width=8cm]{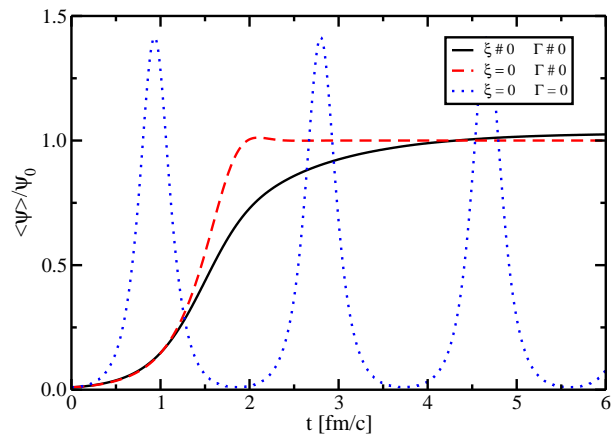}
\caption{Langevin evolution of the $SU(2)$ order parameter.} 
\label{loop_su2}
\end{figure}

The thermal mass $M$ can be determined through the deconfinement temperature. 
For $SU(2)$, $T_d=302$ MeV, so that $M=775$ MeV. In Fig. \ref{loop_su2} 
we show the time evolution of $\langle \psi\rangle $ for the $SU(2)$ case, normalized by
$\psi_0$, which corresponds to the value of the order parameter at the vacuum. The dotted 
line represents the case with no noise and no dissipation, the dashed line corresponds to 
the case with only dissipation, and the full line to the complete case. Simulations were run 
under a temperature of $T=6.6\ T_d$, which ensures that there is no barrier to overcome, 
and the dynamics will be that of spinodal decomposition. For this temperature the value 
of $\Gamma$ is given by $10^3~$fm$^{-2}$, in accordance with the discussion 
of the previous section. 
\vspace{0.5cm}

One can clearly see from the figure that dissipation brings strong effects in the time 
evolution, delaying considerably the necessary time for the onset of the decay 
process. Noise acts in the same direction as dissipation, retarding even more the 
time of equilibration: from around $2$ fm/c, for the simulation including dissipation 
effects only, to more than $4$ fm/c in the complete case. Comparing our results to those 
from a similar calculation performed for the case of the chiral phase 
transition \cite{Fraga:2004hp}, it is evident that in the former dissipation and noise 
have similar but stronger effects. This might signal that the dynamics of the 
deconfinement transition is more sensitive to medium effects. However, this is 
a very premature conjecture, since both effective theory approaches are rather 
simplified descriptions of in-medium QCD.

\section{Conclusions and outlook}

We have presented a systematic procedure to study the real-time dynamics 
of pure gauge deconfinement phase transitions, considering in detail the 
case of $SU(2)$. Given an effective field theory for the order parameter 
of the transition, we have discussed the necessity to introduce counterterms 
from lattice renormalization that guarantee lattice independence of physical 
results. These counterterms were computed for the case of $SU(2)$ or any 
theory whose effective model exhibits the same divergence structure.

For the Langevin evolution, one needs the dissipation coefficient as an input. 
We have described a recipe to extract this kinetic quantity from Glauber dynamics 
in Monte Carlo simulations. The value adopted here is based on preliminary 
lattice results. A detailed analysis will be presented in a future 
publication \cite{next}, together with Langevin evolution results for 
the case of $SU(3)$.

{}From our results for the dynamics of the deconfining transition in $SU(2)$, 
we conclude that dissipation and noise play a very relevant role, being 
responsible for delays in the equilibration time of the order of $100\%$. 
So, effects from the medium are clearly significant in the determination of 
the physical time scales, and should be included in any description.

Of course, the treatment implemented here is very simplified in many respects. 
First, there is a need for a more robust effective theory for the order parameter 
of the deconfining transition. Recently, studies of the renormalization of Polyakov 
loops naturally lead to effective {\it matrix} models for the deconfinement 
transition \cite{matrix}, unfolding a much richer set of possibilities than the approach 
considered here. In particular, eigenvalue repulsion from the Vandermonde determinant 
in the measure seems to play a key role as discussed in Ref. \cite{Pisarski:2006hz}. 
Nevertheless, these studies have shown that, in the neighborhood of the transition, 
the relevant quantity is still the trace of the Polyakov loop. 

Second, there is a need to construct a phenomenological generalized Landau-Ginzburg 
effective theory describing {\it simultaneously} the processes of chiral symmetry restoration 
and deconfinement in the presence of massive quarks as discussed in Ref. 
\cite{Fraga:2007un}. Then, the dynamics of the approximate order parameters, the 
chiral condensate and the expectation value of the trace of the Polyakov loop, will 
be entangled. Finally, if one has the physics of heavy ion collisions in mind, effects 
brought about by the expansion of the plasma \cite{explosive} and by its finite 
size \cite{Fraga:2003mu} will also bring corrections to this picture. 

In a more realistic approach, time scales extracted from the real-time evolution of the 
order parameters can be confronted with high-energy heavy ion collisions experimental 
data, and perhaps provide some clues for the understanding of the mechanism of 
equilibration of the quark-gluon plasma presumably formed at Relativistic Heavy 
Ion Collider (RHIC).

\section*{Acknowledgments} 
We thank G. Ananos, A. Bazavov, A. Dumitru, L. F. Palhares and D. Zschiesche for discussions. 
This work was partially supported by CAPES, CNPq, FAPERJ, FAPESP and FUJB/UFRJ.


\end{document}